\documentclass{jltp}

\usepackage{graphicx}

\title{Influence of Temperature Gradients on Tunnel Junction Thermometry below 1 K: Cooling and %%@
Electron-Phonon Coupling}

\author{J.T. Karvonen, L.J. Taskinen and I.J. Maasilta}

\address{Nanoscience Center, Department of Physics, P. O. Box 35, \\ FIN-40014 University of %%@
Jyv\"askyl\"a, Finland}

\runninghead{J.T. Karvonen, L.J. Taskinen and I.J. Maasilta}{Influence of Temperature Gradients on %%@
Tunnel Junction Thermometry below 1 K}

\begin{document}

\maketitle

\begin{abstract}
We have studied thermal gradients in thin Cu and AlMn wires, both experimentally and %%@
theoretically. In the experiments, the wires were Joule heated non-uniformly at sub-Kelvin %%@
temperatures, and the resulting temperature gradients  were measured using normal %%@
metal-insulator-superconducting tunnel junctions. The data clearly shows that even in reasonably %%@
well conducting thin wires with a short  ($\sim 10$ $\mu$m) non-heated portion, significant %%@
temperature differences can form. In most cases, the measurements agree well with a model which %%@
includes electron-phonon interaction and electronic thermal conductivity by the Wiedemann-Franz %%@
law.

\end{abstract}

\section{INTRODUCTION}
\label{intro}
Temperature is naturally the most critical quantity in studies of any thermal properties of %%@
materials, and its measurement is typically a non-trivial task. In nano- and mesoscopic structures %%@
at low temperatures, this task is made even harder by the small size of the samples and their %%@
sensitivity to any external noise power, which is caused by weakness of the electron-phonon %%@
interaction.\cite{gantmakher} Because of this weakness, electrons can be easily overheated (or %%@
cooled) with respect to phonons at temperatures below 1K, a phenomenon known as the hot-electron %%@
effect.\cite{roukes} In this quasiequilibrium regime, valid in many situations, the electronic and %%@
phononic degrees of freedom attain their own internal equilibrium temperatures, with a power %%@
flowing between them. Given this regime, it is possible to measure the electron temperature, if a %%@
suitable thermometer is found.        
 
 A good choice for electron thermometry below 1K is normal metal-insulator-superconductor (NIS) %%@
tunnel junction thermometers,\cite{rowell,giaz} known for their sensitivity. They can also be %%@
fabricated to a very small size $\sim 100 \times 100$ nm, so that it is possible to measure local %%@
temperatures. However, there are few studies describing how sample geometry affects temperature %%@
profiles in mesoscopic samples, and consequently how it complicates the interpretation of %%@
temperature measurements.      
 
In this paper, we discuss experimental and theoretical results on temperature gradients in %%@
mesoscopic metal wires with non-uniform Joule heating. The experiments were performed with NIS %%@
junctions, therefore allowing us to measure temperatures at several locations. We observe that %%@
there are significant temperature gradients in copper and aluminum manganese wires, even if the %%@
non-heated portion of the wire is only $\sim 10$ $\mu$m long. The experimental observations can be %%@
explained successfully by numerical solution of a non-linear differential equation describing the %%@
heat balance, which incorporates the Wiedemann-Franz law and electron-phonon coupling theory. In %%@
addition, we also present numerical results on temperature profiles caused by non-uniform cooling, %%@
modeling e.g. tunnel junction coolers.\cite{giaz}     

The paper is organized as follows: Sec. 2 discusses the theory. In Subs. 2.1 we briefly review the %%@
theory of electron-phonon interaction in disordered films, followed by the numerical results on %%@
non-uniform Joule heating in 2.2 and on cooling in 2.3. Section 3 presents the experimental %%@
techniques, with the experimental results presented in Sec. 4. Conclusions are drawn in Sec. 5.
%-------------------------------------------------------------------------------------------------%%@
---
\section{THEORY AND NUMERICAL RESULTS}
\label{theory}
\subsection{Electron-phonon coupling in metals}
\label{e-p}
Electron-phonon (e-p) scattering is a critical process for understanding how electron gas relaxes %%@
energy. It is the dominant mechanism for electron energy loss below 1 K, as photon radiative %%@
losses are very small except in the smallest samples.\cite{dan,meschke} The strength of e-p  %%@
coupling depends significantly on several factors: the material in question; the level of disorder %%@
in the metal, parametrized by the electron mean free path $l$;\cite{schmid,sergeev2} and the type %%@
of scattering potential.\cite{sergeev} 

In general, the electron-phonon scattering rate has a form
\begin{equation}
\label{tau}
\frac{1}{\tau_{e-p}}=\alpha T_{e}^{m},
\end{equation} 
and the corresponding net power density transferred from hot electrons to phonons is
\begin{equation}
\label{power}
p=\Sigma (T_{e}^{n}-T_{p}^{n}),
\end{equation}   
where $\Sigma$ describes the strength of electron-phonon coupling, $T_e$ is the electron and $T_p$ %%@
the phonon temperature, and $n=m+2$. The exact form of coupling constant $\Sigma$ and the exponent %%@
$n$ is determined by the disorder, mainly depending on the parameter $ql$, where $q$ is the %%@
wavevector of the dominant thermal phonons. 

In ordered metals, defined as $ql > 1$, electrons can scatter either from longitudinal only, or %%@
from longitudinal and transverse phonons depending on temperature and material.\cite{reiz} When %%@
scattering happens only from longitudinal phonons, the temperature dependence for scattering rate %%@
$\tau_{e-p}^{-1}$ is $m=3$ and, for the heat flow in Eq. (\ref{power}), %%@
$n=5$.\cite{gantmakher,wellstood} In this case, the coupling constant $\Sigma$ is only a material %%@
dependent parameter. If electrons interact dominantly with transverse phonons, $m=2$, $n=4$ and %%@
$\Sigma\propto 1/l$.\cite{sergeev} However, if the scattering rates for transverse and %%@
longitudinal phonons are are of the same magnitude, $m$ can vary continuously between $m=2-3$, so %%@
that $n=4-5$ and $\Sigma \propto l^{-1} - l^{0}$. \cite{sergeev} 

In disordered metals, where $ql < 1$, electrons scatter strongly from impurities, defects and %%@
boundaries, and the situation is more complicated to model because of the interference processes %%@
between pure electron-phonon and  electron-impurity scattering events. However, there is a theory %%@
that includes electron-impurity scattering by vibrating and static disorder.\cite{sergeev} In the %%@
case of fully vibrating impurities (following the phonon mode) $m=4$, $n=6$ and $\Sigma \propto %%@
l$. On the other hand, if the scattering potential is completely static, electrons interact only %%@
with longitudinal phonons and at low temperatures $m=2$, $n=4$ and $\Sigma \propto 1/l$. Between %%@
these two extremes there is a region, where the scattering potential is a mixture of the two, and %%@
theory predicts exponents $m$ ranging between 2-4 and $\Sigma$ depending on $l$ as $l^{k}$, where %%@
$k$ ranges between -1 and 1. 

\subsection{Numerical results on Joule heating}
\label{simulations}
Non-uniform heating and/or cooling profiles generate temperature gradients, even in good %%@
conductors such as copper. The magnitude of the effect is determined by the ratio of the energy %%@
flow due to electronic diffusion (thermal conductance) to that of the energy flow to the phonons. %%@
If diffusive flow is much larger, small temperature gradients exist and vice versa. Therefore, it %%@
is intuitively clear that more resistive samples have larger gradients, if the electron-phonon %%@
interaction does not depend on the mean free path $l$. However, as we discussed in the previous %%@
section, in impure metal films the strength of the e-p interaction does actually depend on $l$, %%@
and {\it a priori} it is not fully clear how this affects the temperature profiles.

To study the temperature profiles in mesoscopic metallic samples, we need to solve numerically the %%@
non-linear differential equation describing the heat flow in the system. We restrict ourselves to %%@
the simplest one-dimensional problem, which is valid for wires of approximately constant cross %%@
section. This is a good approximation for the samples described in the next section (Table %%@
\ref{mitat}). The heat equation for a sample with uniform resistivity $\rho$ reads

\begin{equation}
\label{diffis}
\frac{d}{dx}\left (\frac{{\cal L}}{\rho} T(x) \frac{dT(x)}{dx} \right) = \Sigma %%@
[T(x)^n-T_p^n]-\dot{q}_h(x)+\dot{q}_c(x),
\end{equation}    
where we have used the Wiedemann-Franz law for electronic thermal conductivity $\kappa={\cal %%@
L}T/\rho$ with ${\cal L}=\pi^2k_B^2/(3e^2)$ the Lorentz number,\cite{berman} and where %%@
$\dot{q}_h(x)$ and $\dot{q}_c(x)$ are the power density profiles for heating and cooling, %%@
respectively. If one heats the wire with a dc current density $i$, the Joule heating power density %%@
will be given by $\dot{q}_h(x)=\rho i^2 f(x)$, where $f(x)=1$ at points where current flows, and %%@
$f(x)=0$ elsewhere. Typically, this problem has been solved for Dirichlet boundary conditions, for %%@
which temperature is fixed at the boundary. The Dirichlet problem describes well the case in which %%@
the wire is in direct contact with thick and wide normal electrodes,\cite{martinis} whose %%@
temperature stays constant. In contrast, our sample geometry is such that the wire does not have %%@
any contacts at the physical ends, and the heating current is passed through superconducting leads %%@
in direct contact with the wire (NS boundaries), so that no heat flow takes place through them. In %%@
this case, the correct boundary conditions are the von Neumann type, where $dT/dx=0$. 

\begin{figure}[t]
\centering
\includegraphics[width=0.9\linewidth]{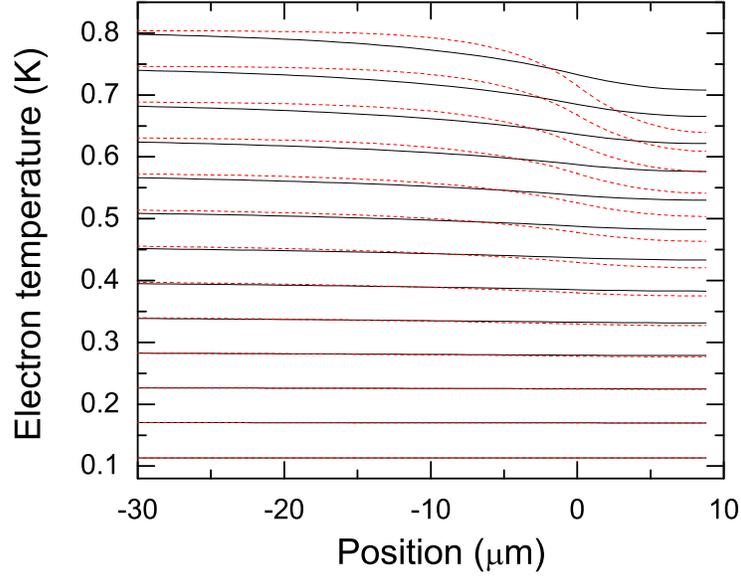}
\caption{Color online. Calculated temperature profiles with varying Joule power levels applied %%@
uniformly at positions $x < 0$. Solid (black) line uses material parameters for Cu sample 2 and %%@
dashed (red) line for the AlMn sample, see Tables \ref{mitat} and \ref{sigmat}. $T_{p}= 60$ mK. }
\label{simu}       
\end{figure}

First, we discuss the numerical results on Joule heated wires. In this case the heated portion of %%@
the wire is much longer than the typical electron-electron scattering length at sub-Kelvin %%@
temperatures ($ L \sim 0.5$ mm $>> L_{e-e} \sim 1$ $\mu$m), and therefore quasiequilibrium with a %%@
well defined electron temperature exists everywhere in the wire. In addition to the heated part of %%@
the wire, a short stub of length $d= 9$ $\mu$m (corresponding to Cu sample 2, see Table %%@
\ref{mitat}) extends beyond the heated portion, where electrons can only diffuse and be cooled by %%@
phonons. This means that the temperature will start to fall, and the magnitude of this drop is a %%@
strong function of temperature itself. The first (and wrong, as we will see) intuition is that if %%@
the stub is much shorter than the length scale for phononic energy relaxation %%@
$L_{e-p}=\sqrt{D\tau_{e-p}}$, where $D$ is the diffusion constant, there is no significant effect %%@
below 1 K, as $L_{e-p}$ ranges between $\sim$1 mm and 10 $\mu$m at 100 mK - 1 K for typical thin %%@
films. 

Figure \ref{simu} shows the calculated temperature profiles for different Joule powers using the %%@
materials parameters for our experimental sample Cu 2 (solid) and the AlMn sample (dashed) (Tables %%@
\ref{mitat} and \ref{sigmat}). The phonon temperature $T_{p}$ was set to 60 mK, which is the %%@
refrigerator temperature used in the experiments. The length of the stub was kept constant, using %%@
the value for the Cu 2 sample $d= 9$ $\mu$m. The x-coordinate is such that Joule power is applied %%@
at $x < 0$, and the Joule power levels between Cu and AlMn are adjusted so that the electron %%@
temperatures in the bulk of the wires are equal. It is very clear that a short stub of this length %%@
does, in fact,  have a significant effect on the profiles, contrary to our initial expectation. At %%@
$T > 300 $ mK, the temperature drop seems to be measurable for both materials, and stronger for %%@
AlMn, which has approximately five times higher resistivity, but also a weaker e-p scattering %%@
rate. Interestingly, the temperature drop extends mostly into the area $x < 0$, where Joule heat %%@
is being {\em uniformly} applied. Therefore, the bulk electron temperature, determined solely by %%@
the electron-phonon scattering, can only be measured at $> 40$ $\mu$m distance away from the end %%@
of the wire. 
In the next section we describe the experiments used to study these temperature gradients.             
  
\begin{figure}[t]
\centering
\includegraphics[width=\linewidth]{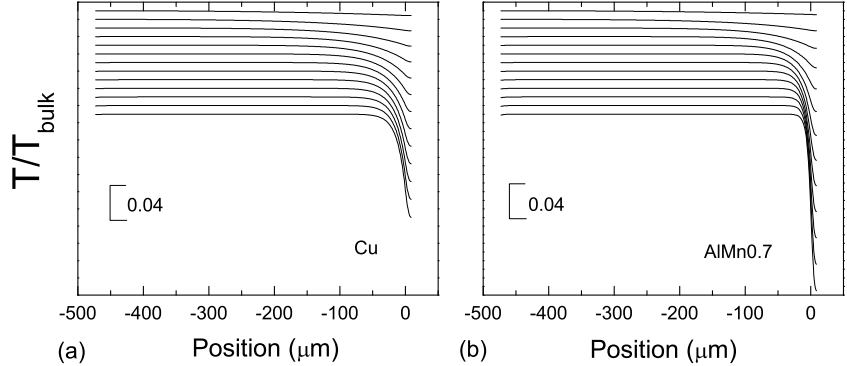}
\caption{Calculated relative temperature profiles $T/T_{bulk}$ with varying Joule power levels %%@
applied at positions $x < 0$. Parameters used are the same as in Fig. \ref{simu}, (a) Cu sample 2, %%@
and (b) AlMn sample. Note that the curves have been shifted for clarity. Top curve corresponds to %%@
the lowest temperature $T_{bulk} \approx 100$ mK. }
\label{simu2}       
\end{figure}

In Fig. \ref{simu2}, we also plot the same profiles scaled to the bulk values for each Joule %%@
power, so that the full temperature profile is clearly seen for all temperatures. We see that the %%@
length scale for the temperature drop grows strongly to $> 100$ $\mu$m as one lowers the bulk %%@
temperature to 100 mK (top curve), although the relative drop becomes small. Also, the temperature %%@
profile is highly non-symmetric with respect to the average temperature, as expected for a %%@
non-linear system. At the higher temperature range, the AlMn profiles become clearly steeper due %%@
to the differences in $\rho$, $\Sigma$ and $n$.   

\begin{figure}[t]
\centering
\includegraphics[width=\linewidth]{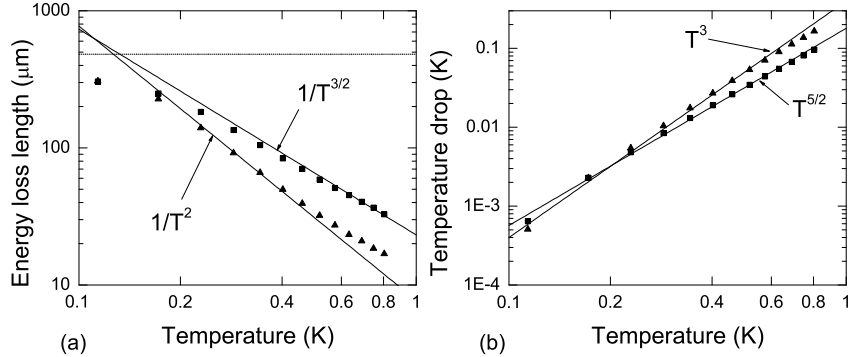}
\caption{(a) Symbols: Calculated energy loss length, defined as the length where $T$ has changed %%@
90 \% of the total change. Squares: Cu, Triangles: AlMn. Solid lines show theoretical values %%@
$2.2L_{e-p}$. Dashed lines shows the length of the wire. (b) Magnitude of the temperature drop, %%@
Squares: Cu, Triangles: AlMn. Solid lines are fits to power laws. Parameters used are the same as %%@
in Fig. \ref{simu}. }
\label{lep}       
\end{figure}

From the calculated profiles, one can see how the length scale and magnitude of the temperature %%@
drop depend on the bulk wire temperature. This is shown in Fig. \ref{lep}. We have defined the %%@
energy loss length as the distance where $T$ changes 90 \% of the total change measured from the %%@
end of the stub. By comparing with the theoretical electron-phonon length $L_{e-p}$ in Fig. 3 (a), %%@
we see that our definition corresponds to roughly $\sim 2.2 L_{e-p}$, with the correct temperature %%@
dependence determined by the e-p scattering $L_{e-p} \sim T^{m/2}$, where $m=3$ for Cu and $m=4$ %%@
for AlMn (Table \ref{sigmat}). The deviations at low temperatures are due to saturation caused by %%@
$2L_{e-p}$ approaching the length of the wire. The high temperature deviation in the case of AlMn %%@
shows that, at higher Joule power levels, the energy loss length is not simply given by $\sim 2 %%@
L_{e-p}$. We do not have a theoretical description for the magnitude of the temperature drop, %%@
which must be a function of the stub length $d$. However, we have simply plotted the values in %%@
Fig. \ref{lep} (b) with the observation that they follow a power law of the form $\Delta T \sim %%@
T^{(m+2)/2}$ for both materials, where $(m+2)$ and 2 are the exponents of the temperature %%@
dependencies of the heat flow rates due to e-p interaction and diffusion, respectively.  

\subsection{Numerical results on NIS tunnel junction cooling}

So far we considered the case in which we had a non-uniform Joule heating profile. Technologically %%@
important is also the case in which a non-uniform cooling profile exists. This is in practice the %%@
case always for normal metal-insulator superconductor (NIS) tunnel junction electron %%@
coolers,\cite{giaz} as they never cover the whole area of the normal metal being cooled (there has %%@
to be room for thermometers). However, if the {\em total} dimensions of the normal metal island %%@
are small compared to $L_{e-p}$ (regardless of how much area is uncooled), no significant %%@
gradients will develop. On the other hand, especially difficult is the case of phonon membrane %%@
coolers,\cite{clark,arttu} where the cooler junctions are located on the bulk of the wafer and %%@
extend a normal metal cold finger onto a thin insulating membrane. In that case, temperature %%@
gradients have to be taken into account. 

\begin{figure}[t]
\centering
\includegraphics[width=0.9\linewidth]{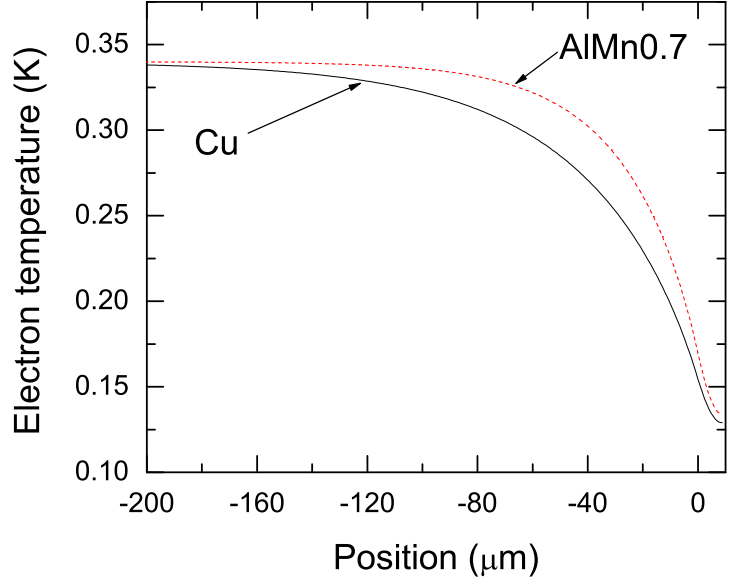}
\caption{Color online. Calculated temperature profiles with a cooling power applied at positions %%@
$x > 0$. Solid (black) is the result for Cu sample 2 and (red) dashed line for the AlMn sample. %%@
Parameters used are the same as in Fig. \ref{simu}. }
\label{cooler}       
\end{figure}

In Fig. \ref{cooler} we plot the calculated temperature profiles for a long wire with a uniform %%@
cooling power applied at positions $x>0$ and for a constant phonon temperature $T_{p}= 340$ mK, %%@
with Cu sample 2 and AlMn sample parameters (Tables \ref{mitat}, \ref{sigmat}). The cooling power %%@
was chosen so that the minimum temperature would reach $\sim $ 100 mK, which is a typical value %%@
for aluminum based tunnel junction coolers. \cite{giaz} It is clear that the temperature starts to %%@
rise already within the cooled area, and rises back to the bulk value of $T_{p}$ within %%@
approximately $100$ $\mu$m, given by $\sim 2 L_{e-p}$ as before. This shows that effective cooling %%@
only works some tens of microns away from the junctions on the bulk substrate. When modeling %%@
membrane coolers more accurately,  it is necessary to take into account the gradients of the %%@
phonon temperature \cite{ullom} and the fact that the electron-phonon interaction strength is %%@
weakened on membranes.            

%-------------------------------------------------------------------------------------------------%%@
---
\section{EXPERIMENTAL TECHNIQUES AND SAMPLES}
\label{samples}
\begin{figure}[t]
\centering
\includegraphics[width=1\linewidth]{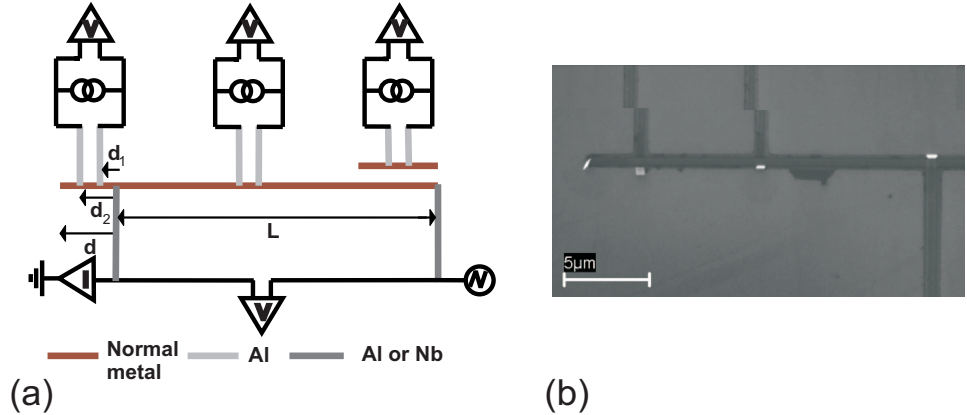}
\caption{(a) Schematic of the sample geometry and the measurement circuit. Black lines are the %%@
normal metal (Cu or AlMn), dark gray Al, and light gray Al or Nb. All the Al leads above the %%@
normal metal wire form NIS tunnel junctions, whereas the Al or Nb leads below have direct NS %%@
contacts. $L$ is the length of the heated part of the wire and $d$ is the length of the unheated %%@
part. $d_{1}$ and $d_{2}$ are the positions of the  NIS-junctions from the nearest NS-interface. %%@
(b) An SEM picture of an SN interface and a SINIS junction at the end of the heated wire of a %%@
representative sample.}
\label{sample}       
\end{figure}

We performed non-uniform heating experiments on several Cu and AlMn wires.  All samples were %%@
fabricated on oxidized or nitridized silicon chips by standard e-beam lithography and multi-angle %%@
shadow mask evaporation techniques. Figure \ref{sample} shows the schematic picture of the samples %%@
and the measurement circuit. Table \ref{mitat} presents the essential dimensions of the samples %%@
measured by SEM and AFM. The resistivity $\rho$ was determined from the I-V measurement of the %%@
wire at 60 mK, from which the mean free path $l$ was calculated using the Drude formula.    

All samples have a normal metal wire of length $\sim$ 500 $\mu$m and width $\sim$ 400-600 nm, onto %%@
which two Al (Nb for Cu sample 2) leads form direct normal metal-superconductor (NS) interfaces. %%@
In addition, two pairs of Cu-AlO$_{x}$-Al (NIS) tunnel junctions connect to the wire. The NS %%@
contacts are used to pass the heating current, and the SINIS junctions serve as electron %%@
thermometers in the middle and at the end of the wire. Near the heated wire, there is also a %%@
short, electrically isolated normal metal wire with a SINIS thermometer, which  measures the local %%@
phonon temperature.    

\begin{table}[h!]
\caption{Parameters of measured Cu and AlMn samples. $L$, $d$, $d_1$ and $d_2$ are defined in Fig. %%@
\ref{sample}, $t$ is the thickness and $A$ the cross sectional area of the normal metal wire. %%@
$\rho$ is the measured resistivity, and $l$ the mean free path.}
\label{mitat} 
\begin{tabular}{lllll}
\hline\noalign{\smallskip}     
Parameter & Cu sample 1& Cu sample 2 & Cu sample 3 & AlMn sample \\
\noalign{\smallskip}\hline\noalign{\smallskip}
 $L$ [$\mu$m] & 473 & 473 & 492 & 466\\
 $d$  [$\mu$m] & 11 & 9& 20 & 16\\
 $d_{1}$ [$\mu$m] & 3 & 2 & 7 & 7 \\
 $d_{2}$ [$\mu$m]& 8 & 8 & 14 & 10\\
 $t$  [nm] & 48 & 32 & 28& 55\\
 $A$ [x10$^{-14}$m$^{2}$]& 1.5& 1.5& 2.5& 1.4\\
 $\rho$  [x10$^{-8}\Omega$m] & 2.5 & 3.0 & 3.8 & 12.3\\
 $l$ [nm]  & 27 & 22 & 17 & 3.2\\
\noalign{\smallskip}\hline
\end{tabular}
\end{table}

Because of Andreev reflection at the NS-junctions, which are biased within the superconducting gap %%@
$\Delta$, the Joule heating current in the normal metal is converted into a supercurrent in the %%@
superconductor, which does not carry any heat with it. Thus, the NS contacts are very good %%@
electrical conductors and, in the ideal case, perfect thermal insulators. This way the Joule heat %%@
does not leak into the superconducting side and the NS contacts do not cause any thermal %%@
gradients. In other words, the Joule heat is uniform between the the NS contacts. While the Joule %%@
current is being applied, we can simultaneously measure the applied power $P$ from the measurement %%@
of current and voltage in four probe configuration, the electron temperature in the middle of the %%@
wire, where no gradients exist, as well as at the end of the wire, where significantly lower %%@
temperatures are expected. More details of the typical thermometer biasing and calibration are %%@
described in Ref. 17. The effect of the thermometers on the temperature profile was estimated to %%@
be insignificant. 

%-------------------------------------------------------------------------------------------------%%@
---
\section{ EXPERIMENTAL RESULTS}
\label{results}
 As the thermometer approximately in the middle of the wire (the middle thermometer) measures the %%@
temperature in a region without thermal gradients (Sec. 2.2), we see that Eq. (\ref{diffis}) will %%@
reduce to $\dot{q}_h=\Sigma(T_e^n-T_p^n)$, and the values of $n$ and $\Sigma$ can be determined %%@
from the data. The thermometer at the unheated end of the wire (the side thermometer), on the %%@
other hand, lies in the region of a gradient. As it comprises of two NIS-tunnel junctions %%@
separated by a small, but a significant distance, the two junctions measure two different %%@
temperatures, and the combined SINIS measurement will give a value in between these two %%@
temperatures. The measured temperature is not necessarily the average of the two due to the %%@
nonlinearity of the NIS thermometer.  

\begin{figure}
\includegraphics{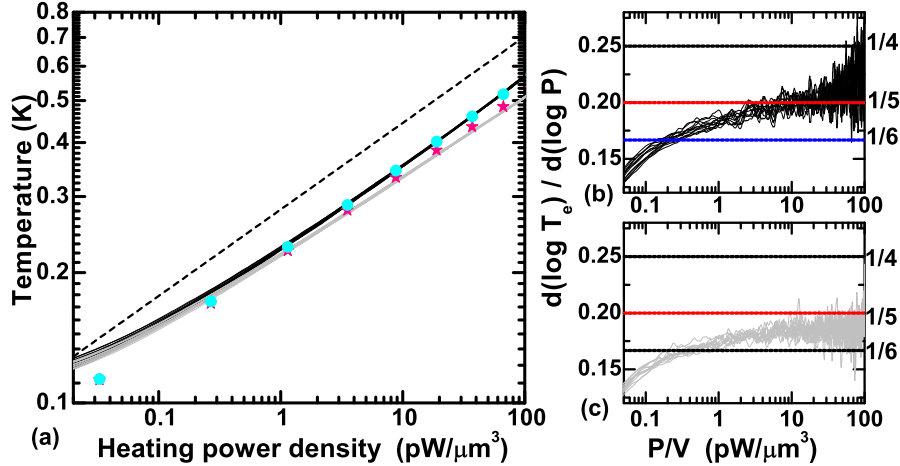}
\caption{Color online. Data from Cu sample 1. (a) The temperatures of the thermometers vs heating %%@
power density in log-log scale.  Black line: experimental data from the middle thermometer. Gray %%@
line: experimental data from the side thermometer. Gray (Cyan) circles: numerical results for the %%@
middle thermometer, and gray (pink) stars: numerical results for the side thermometer. Dashed line %%@
is a guide to the eye  $T \propto$ $(P/V)^{1/5}$. (b) The logarithmic numerical derivatives of the %%@
experimental middle thermometer data, and (c) the same for the side thermometer.} 
\label{Cusample1}      
\end{figure}

\begin{figure}
\includegraphics{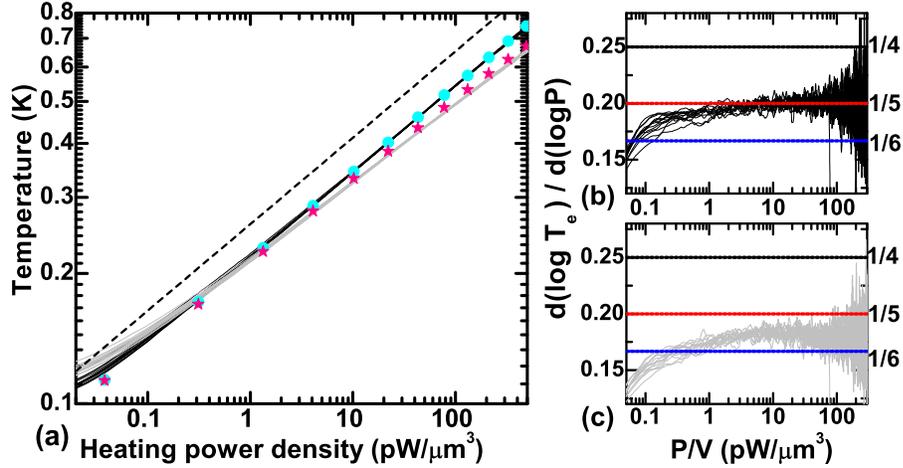}
\caption{ Data from Cu sample 2. Symbols are explained in the caption of  \ref{Cusample1}.}
\label{Cusample2}      
\end{figure}

Cu samples 1 and 2 have a similar sample geometry, except for a difference in the thickness $t$, %%@
and in the electron mean free path $l$. In Fig.s \ref{Cusample1}(a) and \ref{Cusample2}(a) we plot %%@
the temperatures of both thermometers vs heating power density $P/V$ in log-log scale. Clear %%@
temperature difference between the two thermometers can be seen at $P/V > 1$ pW/$\mu$m$^3$, and at %%@
100 pW/$\mu$m$^3$ the difference is roughly 50 mK. Below 0.1 pW/$\mu$m$^3$ both temperatures %%@
saturate mostly because of noise heating. The theoretical points have been calculated by solving %%@
Eq. (\ref{diffis}) using the appropriate sample parameters. In the calculation, we used values for %%@
$\Sigma$ obtained from the middle thermometer data (Table \ref{sigmat}). The temperature of the %%@
side SINIS thermometer is defined in our analysis as the average of the two NIS-junction %%@
temperatures. This definition is justified, because the difference between the two temperatures is %%@
small and within the size of the plotted datapoints for all sample geometries studied in this %%@
paper. From Figs.s \ref{Cusample1}(a) and \ref{Cusample2}(a) it is clear that the theory agrees %%@
well with the experimental data, showing that the observed temperature difference is fully %%@
explained by phonon cooling and diffusion.   

From the phonon thermometer data (not shown), we have seen that $T_{e} \gg T_{p}$ for all samples %%@
in this work, and therefore for the middle thermometer we can approximate $P_{heat}/V=\Sigma %%@
T_{e}^{n}$, where $V$ is the volume of the heated portion of the wire.  The temperature dependence %%@
$n$ and strength $\Sigma$ of electron-phonon interaction can thus be obtained from the plots of %%@
the middle thermometer temperature versus heating power density in s \ref{Cusample1}(a) and %%@
\ref{Cusample2}(a). A more detailed look at $n$ can be obtained by plotting the logarithmic %%@
derivatives $d(\log T_{e}) / d(\log P)=1/n$, shown in Figs.s \ref{Cusample1}(b) and %%@
\ref{Cusample2}(b). Measured data in both samples scales clearly as $P \propto T_{e}^{5}$. %%@
However, because of the temperature drop at the end of the wire, the data from the side %%@
thermometer, Figs. \ref{Cusample1}(c) and \ref{Cusample2}(c), show a temperature dependence $P %%@
\propto T_{e}^{5.5}$. This exponent does not correspond to the actual power law of the e-p %%@
interaction. 

\begin{figure}[t]
\includegraphics{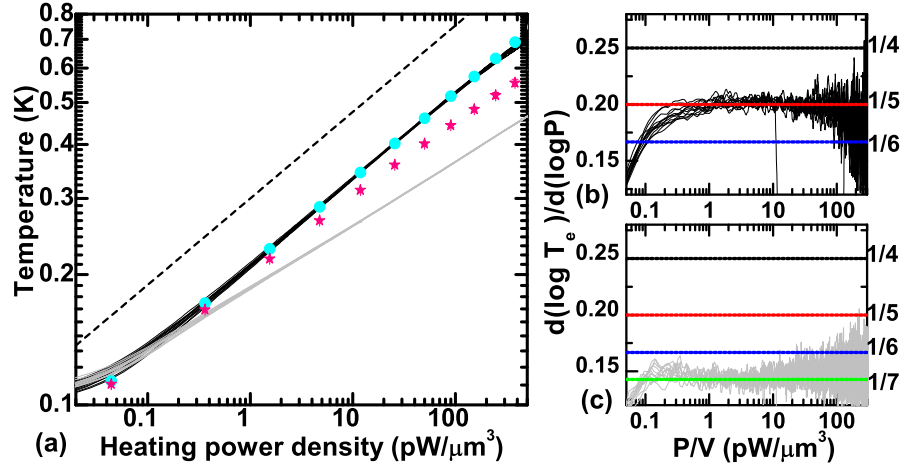}
\caption{Data from Cu sample 3. Symbols are explained in the caption of Fig. \ref{Cusample1}.}
\label{Cusample3}      
\end{figure}

Copper sample 3 is a bit thinner and has a longer unheated end section compared to samples 1 and 2 %%@
(Table \ref{mitat}). From Fig. \ref{Cusample3}(a) we observe that the measured temperature %%@
difference between the two thermometers is much larger than for samples 1 and 2, and also larger %%@
than what the theoretical calculation predicts.  We do not fully understand this at the moment, %%@
but it is possible that the Lorentz number is reduced due to inelastic scattering on the %%@
surface.\cite{berman} Indeed, high resolution SEM images of sample 3 showed that the surface was %%@
more irregular. Nevertheless, we can obtain the temperature dependence of e-p interaction from the %%@
middle thermometer [Figs. \ref{Cusample3}(a), (b)], showing again an agreement with $P \propto %%@
T_{e}^{5}$, while the data from the side thermometer fits $P \propto T_{e}^{7}$.   
 
\begin{figure}[t]
\includegraphics{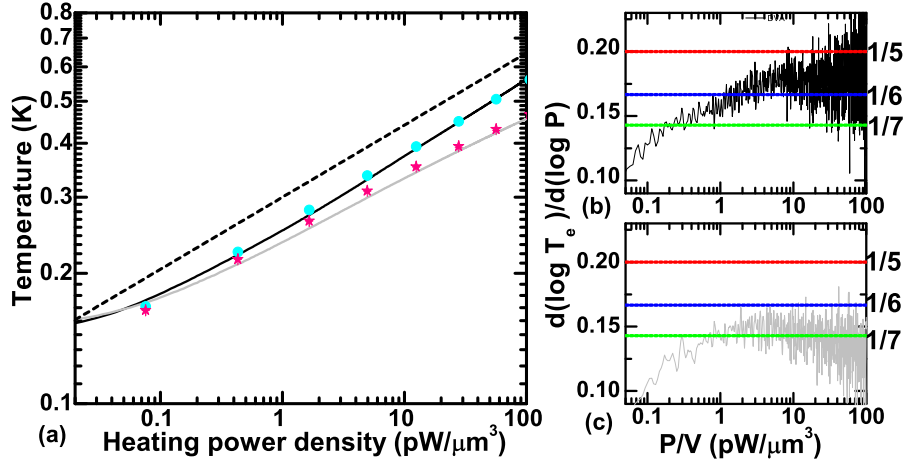}
\caption{Data from the AlMn sample. Symbols are explained in the caption of Fig. \ref{Cusample1}.}
\label{AlMn}      
\end{figure}

The last sample is made from aluminum doped with 0.7 at \% manganese. Due to the high impurity %%@
concentration, the mean free path $l$ is much shorter than for copper wires, and the AlMn wire is %%@
$\sim $ 4 times more resistive (Table \ref{mitat}). Figure \ref{AlMn}(a) shows that the measured %%@
data and the numerical result agree well. It may be surprising, perhaps, that the temperature %%@
difference between the thermometers is actually smaller than in Cu sample 3. Although the AlMn %%@
sample is more resistive and thus diffusion is weaker, the e-p scattering is much weaker. %%@
Therefore, the unheated end of the wire is not as effectively cooled by phonons in AlMn.   

The sample is clearly in the limit $ql \ll$ 1 and the scattering potential is dominated by the Mn %%@
impurities. The data from the middle thermometer, Figs.s \ref{AlMn}(a) and (b) show that $P %%@
\propto T_{e}^{6}$. This is in agreement with the theory including vibrating scatterers, %%@
\cite{sergeev} and has also been observed for other Mn concentrations.\cite{lasse} Again, the side %%@
thermometer gives a much higher apparent temperature dependence: $P \propto T_{e}^{7}$.  

The values for coupling constants $\Sigma$ can be determined from the slopes of the ($ P/V, %%@
T$)-plots in logarithmic scale as fitting parameters, keeping $n$ fixed at $n=5$ or $n=6$. In %%@
Table \ref{sigmat} the measured values for $\Sigma$ are presented for all the samples. For Cu %%@
samples 1-3, $\Sigma$ decreases as a function of electron mean free path $l$; in other words, the %%@
electron-phonon interaction weakens with increased purity of the samples. This is evidence that %%@
the theory for pure electron-phonon coupling does not apply in our Cu thin films, although the %%@
temperature dependence agrees with the simplest theory without disorder.\cite{wellstood} Possible %%@
explanations for the experimental result $P \propto T^5$ are that our Cu samples are either in the %%@
transition region $ql \sim$ 1 or that the scattering potential is not fully vibrating. Our earlier %%@
conclusions\cite{karvonen,solidi} on  the temperature dependence of the electron-phonon coupling %%@
in disordered Cu and Au films were not correct, because the temperatures were measured with a side %%@
thermometer. For aluminum manganese samples with varying impurity concentration, $\Sigma$ is %%@
linearly dependent on $l$, consistent with the fully vibrating disorder theory.\cite{lasse}  
\begin{table}
\caption{Measured values for coupling constant $\Sigma$}
\label{sigmat} 
\begin{tabular}{lllll}
\hline\noalign{\smallskip}     
 & Cu sample 1& Cu sample 2 & Cu sample 3 & AlMn sample \\
\noalign{\smallskip}\hline\noalign{\smallskip}
$n$ & 5 & 5 & 5 & 6 \\
$\Sigma$ [W/K$^n$m$^3$] & 1.8 $\times 10^{9}$ & 2.1 $\times 10^{9}$ & 2.5 $\times 10^{9}$ & 3.4 %%@
$\times 10^{9}$ \\
\end{tabular}
\end{table}

%-------------------------------------------------------------------------------------------------%%@
---
\section{CONCLUSIONS}
\label{conclusions}

We have shown that thermal gradients are easily generated in mesoscopic samples at sub-Kelvin %%@
temperatures, even for good conductors such as copper. This fact has a strong effect on studies of %%@
thermal properties and thermometry. To obtain correct information on the electron-phonon %%@
interaction strength, for example, one has to make sure that electron temperature is measured in a %%@
location without thermal gradients. In addition, our results also imply that for tunnel junction %%@
coolers,  large cold fingers outside the junction area are not effectively cooled. Also, even if %%@
the non-heated or non-cooled area is smaller than the electron-phonon scattering length $L_{e-p}$, %%@
thermal gradients will develop as long as the total size of the normal metal is of the order of %%@
$L_{e-p}$.

%-------------------------------------------------------------------------------------------------%%@
---

\section*{ACKNOWLEDGMENTS}
The authors thank D. E. Prober for valuable discussions. This work was supported by the Academy of %%@
Finland under contracts No. 105258 and 205476.

\end{document}